\documentclass[prl,reprint,twocolumn,showpacs,superscriptaddress]{revtex4-1}
\usepackage{graphicx}
\usepackage{dcolumn}
\usepackage{bm}
\usepackage{graphicx}
\usepackage{epsfig}
\usepackage{epsf}
\usepackage{amssymb}
\usepackage{amsmath,empheq,cases}
\usepackage{amsthm}
\usepackage{mathtools}
\usepackage{multirow}
\usepackage[colorlinks=true,linkcolor=blue,citecolor=blue,urlcolor=blue,pdfauthor={ },pdftitle={ },pdfsubject={ },pdfkeywords={ }]{hyperref}
\usepackage{pgfplots}
\usepackage{lipsum}

\usepackage{times}

\usepackage{tikz}
\definecolor{c2}{RGB}{153,51,51}

\begin{document}

\title{Experimental Observation of Equilibrium and Dynamical Quantum Phase Transitions via Out-of-Time-Ordered Correlators}

\author{Xinfang Nie}
\thanks{These authors contributed equally to this work.}
\affiliation{Shenzhen Institute for Quantum Science and Engineering and Department of Physics, Southern University of Science and Technology, Shenzhen 518055, China}
\affiliation{CAS Key Laboratory of Microscale Magnetic Resonance and Department of Modern Physics, University of Science and Technology of China, Hefei 230026, China}

\author{Bo-Bo Wei}
\thanks{These authors contributed equally to this work.}
\affiliation{School of Science and Engineering, The Chinese University of Hong Kong, Shenzhen, Shenzhen 518172, China}

\author{Xi Chen}
\affiliation{CAS Key Laboratory of Microscale Magnetic Resonance and Department of Modern Physics, University of Science and Technology of China, Hefei 230026, China}

\author{Ze Zhang}
\affiliation{Shenzhen Institute for Quantum Science and Engineering and Department of Physics, Southern University of Science and Technology, Shenzhen 518055, China}
\author{Xiuzhu Zhao}
\affiliation{Shenzhen Institute for Quantum Science and Engineering and Department of Physics, Southern University of Science and Technology, Shenzhen 518055, China}
\author{Chudan Qiu}
\affiliation{Shenzhen Institute for Quantum Science and Engineering and Department of Physics, Southern University of Science and Technology, Shenzhen 518055, China}
\author{Yu Tian}
\affiliation{Shenzhen Institute for Quantum Science and Engineering and Department of Physics, Southern University of Science and Technology, Shenzhen 518055, China}
\author{Yunlan Ji}
\affiliation{Shenzhen Institute for Quantum Science and Engineering and Department of Physics, Southern University of Science and Technology, Shenzhen 518055, China}

\author{Tao Xin}
\email{xint@sustech.edu.cn}
\affiliation{Shenzhen Institute for Quantum Science and Engineering and Department of Physics, Southern University of Science and Technology, Shenzhen 518055, China}

\author{Dawei Lu}
\email{ludw@sustech.edu.cn}
\affiliation{Shenzhen Institute for Quantum Science and Engineering and Department of Physics, Southern University of Science and Technology, Shenzhen 518055, China}
\author{Jun Li}
\email{lij3@sustech.edu.cn}
\affiliation{Shenzhen Institute for Quantum Science and Engineering and Department of Physics, Southern University of Science and Technology, Shenzhen 518055, China}

\date{\today}

\begin{abstract}
The out-of-time-ordered correlators (OTOC) have been established as a fundamental concept for quantifying quantum information scrambling and diagnosing quantum chaotic behavior. Recently, it was theoretically proposed that the OTOC can be used as an order parameter to dynamically detect both equilibrium quantum phase transitions (EQPTs) and dynamical quantum phase transitions (DQPTs) in one-dimensional many-body systems. Here we report the first experimental observation of EQPTs and DQPTs in a quantum spin chain via quench dynamics of OTOC on a nuclear magnetic resonance quantum simulator. We observe that the quench dynamics of both the order parameter and the two-body correlation function cannot detect the DQPTs, but the OTOC can unambiguously detect the DQPTs. Moreover, we demonstrate that the long-time average value of the OTOC in quantum quench signals the equilibrium quantum critical point and ordered quantum phases, thus one can measure the EQPTs from the non-equilibrium quantum quench dynamics. Our experiment paves a way for experimentally investigating DQPTs through OTOCs and for studying the EQPTs through the non-equilibrium quantum quench dynamics with quantum simulators.
\end{abstract}

\maketitle
\emph{Introduction.} --
Equilibrium quantum phase transitions (EQPTs)~\cite{Sachdev2011} are one of the most significant phenomena in many-body physics since it signals new states of quantum matter.
It is accompanied by a nonanalytic change of some physical observable at a quantum critical point and is well understood from the paradigm of renormalization group theory~\cite{Cardy:1996th}. Recently, dynamical quantum phase transitions (DQPTs) that emerge in the dynamics of an isolated quantum many-body systems have attracted extensive theoretical efforts~\cite{PhysRevLett.110.135704,PhysRevLett.113.205701,PhysRevLett.115.140602,PhysRevB.89.125120,PhysRevB.93.085416,PhysRevLett.117.086802,
zvyagin2016dynamical,PhysRevB.93.144306,PhysRevB.95.075143,PhysRevLett.120.130601,heyl2018dynamical,PhysRevA.98.022129} and experimental interests~\cite{zhang2017observation2,PhysRevLett.119.080501,
PhysRevLett.119.080501,PhysRevApplied.11.044080,wang2018simulating,bernien2017probing,flaschner2018observation,PhysRevB.100.024310}. There are two different types of DQPTs. One type is witnessed by the nonanalyticity in the rate function of the Loschmidt echo at critical times~\cite{PhysRevLett.110.135704}, which resembles the nonanalyticity of free energy density as a function of temperature or other control parameters in the EQPTs. The other type is revealed by nonanalyticity of some local order parameters in quench dynamics measured at long time limit as a function of the control parameter of the quenched Hamiltonian~\cite{PhysRevLett.120.130601}.
Both types of DQPTs are intrinsically dynamical quantum phenomena without equilibrium counterparts~\cite{heyl2018dynamical}.

EQPTs and DQPTs are both connected to the large quantum fluctuations~\cite{Sachdev2011,Cardy:1996th} and therefore related to fast propagation of quantum information in many-body systems, which can be captured by a recently proposed out-of-time-ordered  correlations (OTOC)~\cite{shenker2014black,shenker2014multiple}. OTOC is defined as
\begin{equation}\label{Eq0}
O(t)=\langle W(t)^\dagger V^\dagger W(t) V\rangle
\end{equation}
for a given physical system described by a Hamiltonian $H$ and an initial state $\vert \psi_0\rangle$.
Here, $W$ and $V$ are two local Hermitian operators, where $W(t)=U^\dagger(t) W(0)U(t)$ is an operator in the Heisenberg picture with time evolution operator $U(t)=e^{-iHt}$,
and $\langle\cdot\rangle$ denotes the expectation value over the initial state $\vert \psi_0\rangle$. OTOC has been proposed to describe dispersions of local quantum information in quantum many-body
systems, termed information scrambling~\cite{chen2016universal,banerjee2017solvable,he2017characterizing,shen2017out,slagle2017out,fan2017out,huang2017out,iyoda2018scrambling,lin2018out,pappalardi2018scrambling,zhang2019information}. Moreover, it has triggered numerous applications in far-from-equilibrium quantum phenomena,
ranging from nonequilibrium statistical mechanics~\cite{PhysRevA.95.012120},
quantum thermalization~\cite{eisert2015quantum,Bohrdt_2017,swingle2018unscrambling,PhysRevE.95.062127} to black holes~\cite{PhysRevLett.120.231601,magan2018black}.
A recent theoretical study~\cite{heyl2018detecting} proposes that quench dynamics of the OTOC can be
used to detect EQPTs and DQPTs in many-body systems~\cite{PhysRevLett.123.140602,sun2018out,PhysRevB.100.195107}. However, experimental progress on
this OTOC-based detection scheme has been elusive.
%

Here, we report the first experimental observation of the EQPTs and DQPTs from the quench dynamics of OTOC.
Specifically, we simulate the quantum Ising models, including an integrable model and a nonintegrable model, on a four-qubit quantum simulator with the nuclear magnetic resonance (NMR) technique. We measure the quench dynamics of both the two-body correlation function of the order parameter and the OTOC of the order parameter experimentally from a fully polarized initial quantum state. On the one hand, we observe that the two-body correlation function cannot signal the DQPTs but the OTOC can clearly detect the DQPTs both in an integrable and non-integrable quantum Ising models, which experimentally establishes OTOCs as a useful probe of DQPTs. On the other hand, we experimentally show that the long-time average value of the OTOC in quench dynamics signals the equilibrium critical points in both integrable and nonintegrable quantum models, showing that one can extract the equilibrium quantum critical properties from the non-equilibrium quantum quench dynamics.

\emph{Quenches in Ising models.} --
\begin{figure}
  \centering
  \includegraphics[width=0.48\textwidth]{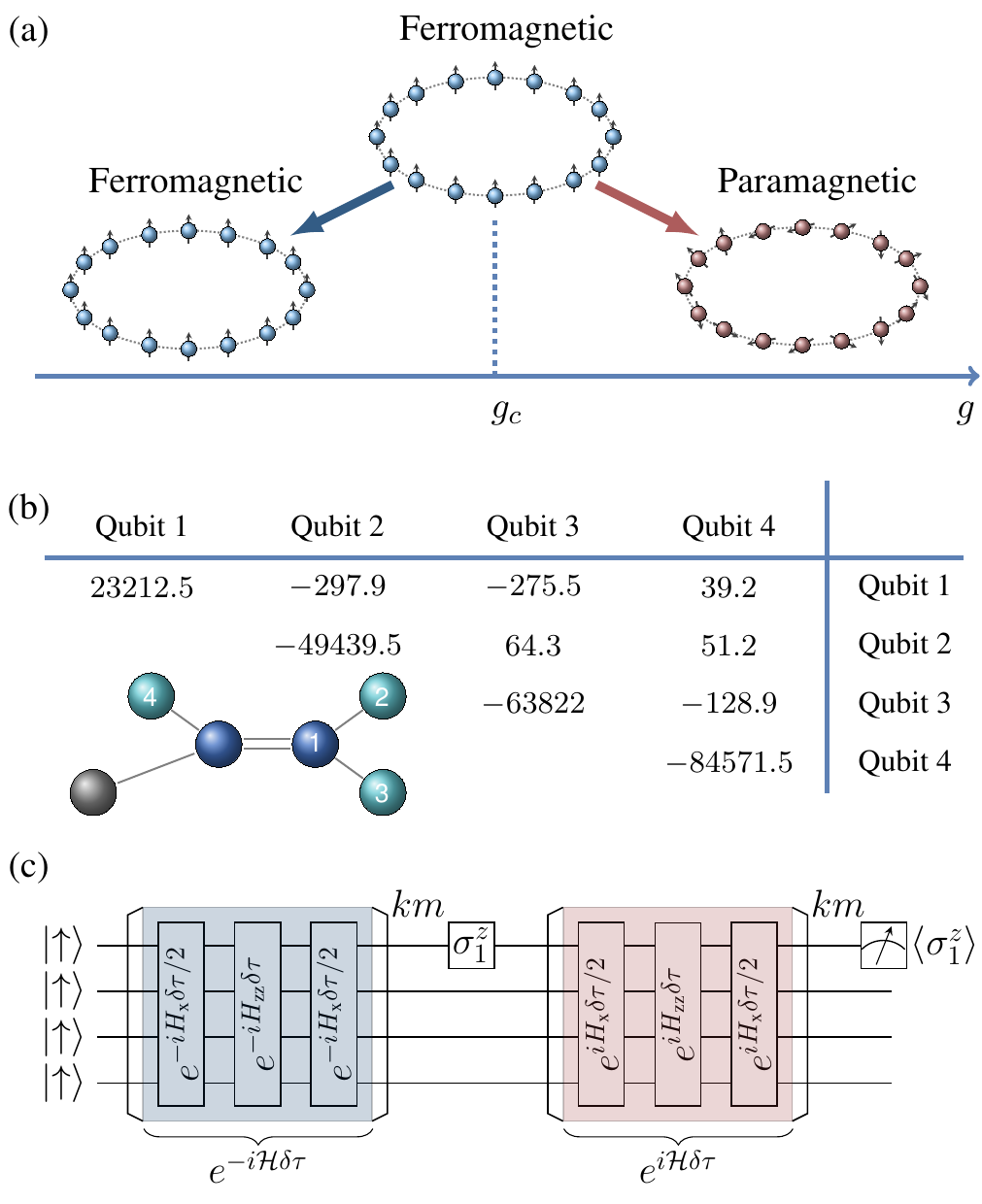}\\
  \caption{
  (a) Illustration of the two kinds of sudden quantum quenches  in a periodic one-dimensional ferromagnetic Ising chain from initial ferromagnetic phase to  (i)  ferromagnetic phase  ($g<g_c$) , and (ii)  paramagnetic phase ($g>g_c$).
  (b) Molecular structure and the Hamiltonian parameters of $^{13}$C-iodotrifluoroethylene (C$_2$F$_3$I).
  The precession frequencies $\omega_i$ and the scalar coupling strengths are given by the diagonal and off-diagonal elements in the table respectively (in units of Hz).
  (c) Quantum circuit diagram of the experiment to detect the OTOC $F(t)$ of the one-dimensional Ising chains.}
  \label{Fig1}
\end{figure}
We study the quench dynamics of ferromagnetic one-dimensional transverse-field Ising model with periodic boundary condition,
as shown in Fig.~\ref{Fig1}(a). The corresponding Hamiltonian is written as
\begin{equation}\label{Eq1}
H=-\sum_{n=1}^{N}[J\sigma^z_n\sigma^z_{n+1}+\Delta\sigma^z_n\sigma^z_{n+2}+g \sigma^x_n],
\end{equation}
where $\sigma^\alpha_n$ $ (\alpha=x, y, z)$ are Pauli operators on the $n$-th site,
$J$ and $\Delta$ denote nearest-neighbor (NN) and the next-nearest-neighbor (NNN) couplings,
and $g$ is the uniform transverse field. For a ferromagnetic Ising model ($J>0$), we assume $J=1$ without loss of generality.

We investigate two kinds of Ising chains:
the integrable version with only nearest interactions, i.e., $\Delta=0$, which is termed as transverse-field Ising chain (TFIC),
and the nonintegrable one with both NN and NNN interactions, namely axial next-nearest-neighbor Ising model (ANNNI).
Both models serve as paradigms in EQPT
~\cite{Sachdev2011,PhysRevB.87.195104,bernien2017probing,PhysRevB.93.144306,SELKE1988213,chakrabarti2008quantum}.
The TFIC undergoes a quantum phase transition at the critical point $g_c=1$:
it stays in the ferromagnetic state for $g<1$,
and in the paramagnetic phase for $g>1$.
The quantum phase diagram is much more complex for the ANNNI~\cite{SELKE1988213,chakrabarti2008quantum,PhysRevB.24.6620,Peschel1981,PhysRevB.73.052402,PhysRevB.87.195104}.
Here, we consider the phase transition between ferromagnetic and paramagnetic phases in the case of $\Delta=0.5$,
where the critical point locates at $g_c\simeq1.6$.
The DQPTs in such Ising models have been theoretically and experimentally studied by Loschmidt echoes~\cite{PhysRevB.87.195104,PhysRevLett.110.135704,PhysRevLett.119.080501,zhang2017observation2}.

\begin{figure*}
  \centering
  \includegraphics[width=\textwidth]{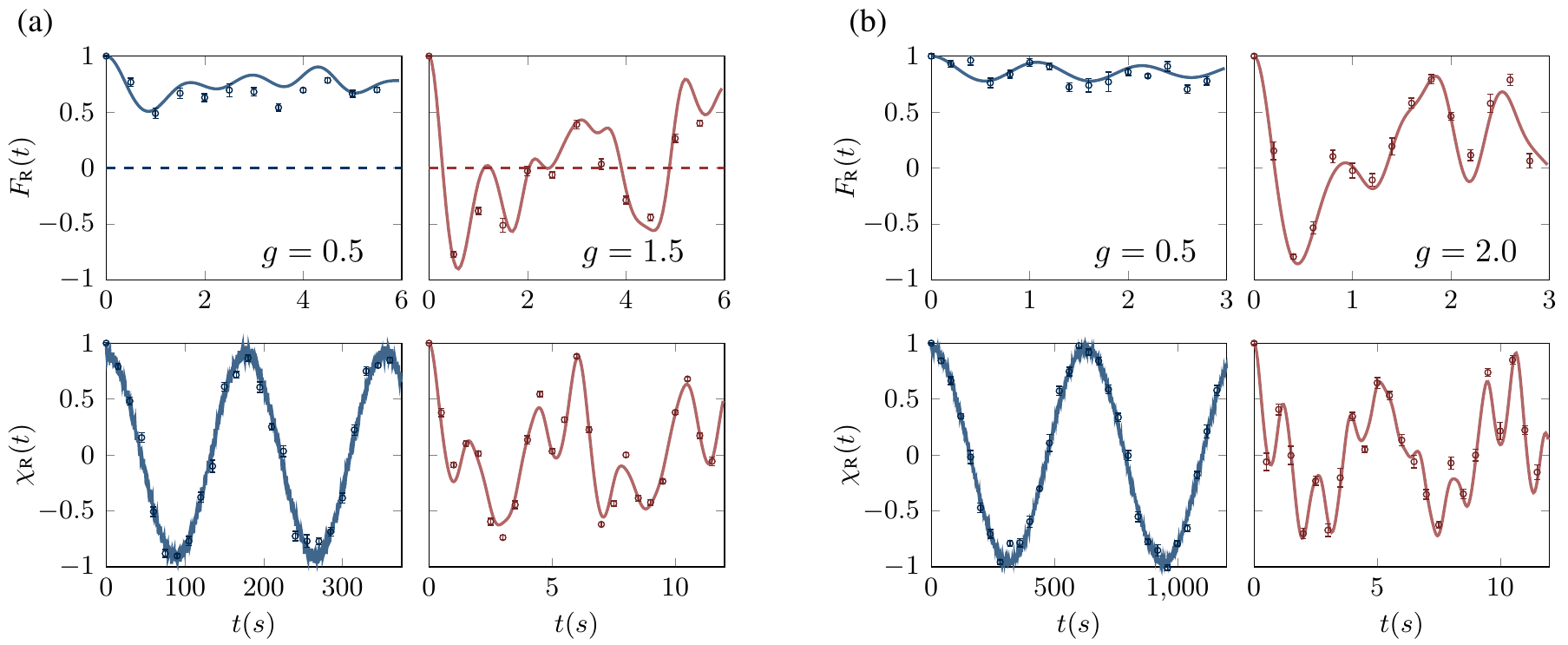}\\
  \caption{Experimentally measured real parts of OTOC $F_R(t)$ (top panels) and autocorrelation $\chi_R(t)$ (bottom panels) of the quantum quench dynamics
   as a function of time $t$
  in (a) TFIC model with $\Delta=0$ and (b) ANNNI model with $\Delta=0.5$.
  Both systems start from the fully polarized state $|\psi_0\rangle$ and then undergo unitary evolutions governed by the Hamiltonian $H(g)$.
  The dots are the experimental data, the solid lines are the numerical simulation results,
  and the error bars are computed from the imperfections of the pulses.}
  \label{Fig2}
\end{figure*}

DQPTs are usually investigated by quantum quenches.
It starts from the ground state of an initial Hamiltonian $H_0$,
and then evolves under another Hamiltonian $H_f$.
For example, in the Ising model in Eq.~\eqref{Eq1}, we choose $\vert\psi_0\rangle$ as
the fully polarized state $\left|  \uparrow  \uparrow  \uparrow\ldots \right\rangle$,
which is one of the two degenerated ground states of initial Hamiltonian with $g=0$.
There are two kinds of quantum quenches in the transverse-field Ising model as shown in Fig.~\ref{Fig1}(a),
depending on the case
$g < g_c$ or $g > g_c$.
DQPT only occurs in the second case when the system quenches across $g_c$.

The autocorrelation function $\chi(t)=\langle \sigma_n^z(t)\sigma_n^z\rangle$~\cite{Sachdev2011},
which can detect equilibrium dynamics, becomes indistinctive in the nonequilibrium case.
It is proposed that the second moment of the autocorrelation function~\cite{heyl2018detecting},
i.e.,
\begin{equation}\label{Eq2}
  F(t)=\langle \sigma_n^z(t)\sigma_n^z\sigma_n^z(t)\sigma_n^z\rangle,
\end{equation}
can be used to distinguish the two kinds of quench dynamics.
In fact, this function $F(t)$ corresponds to the OTOC $O(t)$ in
Eq.~\eqref{Eq0} when the two local operators $W$ and $V$ are both chosen
to be $\sigma_n^z$.
In experiment, we set $W=V=\sigma_1^z$.
Through observing the time dependence of the real part $F_\text{R}(t)$,
one can obtain the information about whether the time evolving Hamiltonian
$H$ is in the ferromagnetic or paramagnetic region.
Moreover, the time-averaging OTOC $\bar{F}_\text{R}=\frac{1}{t}\int_{\tau=0}^tF_\text{R}(\tau)d\tau$
also serves as an order parameter for DQPT:
$\bar{F}_\text{R}$ is nonzero for ferromagnetic phase and vanishes gradually
upon approaching the critical point, while in contrast it stays zero in the whole paramagnetic phase.


\emph{Experiment.} --
The experiments are carried out on a Bruker Ascend $600$ MHz spectrometer ($14.1$ T) equipped with a cryo probe.
The physical system used to perform quantum simulation is the ensemble of  $^{13}$C-iodotrifluoroethylene (C$_2$F$_3$I) dissolved in $d$-chloroform.
The $^{13}$C nuclear spin (Qubit 1) and the three $^{19}$F nuclear spins (Qubits 2-4) constitute a four-qubit quantum simulator.
Each nuclear spin corresponds to a spin site in the Ising model.
The natural Hamiltonian of the  nuclear system placed in a static magnetic field along $z$-direction is
\begin{equation}\label{Eq3}
  H_\text{NMR}=-\sum_{i=1}^4\frac{\omega_{i}}{2}\sigma^z_i+\sum_{i<j,=1}^4\frac{\pi J_{i,j}}{2}\sigma^z_i\sigma^z_j,
\end{equation}
where $\omega_{i}/2\pi$ is the Larmor frequency of the $i$-th spin, $J_{i,j}$ is the scalar coupling between the $i$-th and $j$-th spin.
The molecular structure and the NMR Hamiltonian parameters of the sample are given in Fig. ~\ref{Fig1} (b).

Because $\vert\psi_0\rangle$ is an eigenstate of $\sigma_1^z$,
the OTOC in Eq.~\eqref{Eq2} can be rewritten as
\begin{equation}\label{Eq4}
 F(t)=\langle \psi(t)\vert  \sigma_1^z \vert \psi(t)\rangle,
\end{equation}
with $\vert \psi(t)\rangle=e^{iHt}\sigma_1^ze^{-iHt}\vert\psi_0\rangle$.
In other words, to measure the OTOC throughout the quench dynamics,
we need to initialize the system to the fully polarized state $\vert \psi_0\rangle$,
then apply a unitary transformation $U(t)=e^{iHt}\sigma_1^ze^{-iHt}$ and
finally measure the expectation value $\langle\sigma_1^z\rangle$ of the final state.
The whole experimental process is illustrated in Fig.~\ref{Fig1} (c).
The major part of the experiment is to simulate the unitary operation $U(t) $, which we describe in the following.


Given the target Hamiltonian $H$,
we divide the quench dynamics into $M$ discrete time steps,
and record the instantaneous OTOCs at time $t=k\tau$ ($k=0,1,\cdots, M-1$).
The unitary evolution $U(t)$ can be decomposed into  a sequence of unitary transformations:
a time evolution operator $e^{-iHk\tau}$,
a single qubit rotation $\sigma_1^z$ and a backward time evolution $e^{iHk\tau}$.
The key point of the unitary evolution lies in the realization of the two operators $e^{-iH\tau}$ and $e^{iH\tau}$.
Utilizing the Trotter-Suzuki decomposition formula, the time evolution $e^{-iH\tau}$ can be simulated approximately by
\begin{equation}\label{Trotter}
  e^{-iH\tau}\approx [e^{-iH_\text{x}\delta\tau/2} e^{-iH_\text{zz}\delta\tau} e^{-iH_\text{x}\delta\tau/2}]^m,
\end{equation}
where the evolution time $\tau$ is divided into $m$ segments with equal time length $\delta\tau=\tau/m$. Here,
$H_\text{x}=-g\sum_{n=1}^{4}\sigma_x^{n}$, $H_\text{zz}=-\sum_{n=1}^4\sigma_z^{n}\sigma_z^{n+1}$ for the TFIC model,
and $H_\text{zz}=-[\sum_{n=1}^4\sigma_z^{n}\sigma_z^{n+1}+\Delta\sum_{n=1}^4\sigma_z^{n}\sigma_z^{n+2}]$ for the ANNNI model.
In each segment of evolution, $e^{-iH_\text{x}\delta\tau/2}$ and $e^{-iH_\text{zz}\delta\tau}$
can be realized through optimized radio-frequency pulses combined with the NMR refocusing technique.
The reverse time evolution $e^{iH\tau}$ can also be done in a similar way.
As to the operator $\sigma_1^z$, it is a $\pi$ rotation about the $z$-axis on the first nucleus.
In experiment, to improve the control accuracy, we engineer the unitary evolution $U(t)$ with a shaped pulse optimized by the
gradient ascent technique~\cite{khaneja2005optimal}.
The width of the shaped pulse for each $U(t)$ is $40$ ms with theoretical fidelity above $99.5\%$.

\emph{Integrable TFIC model.} --
We first study the quench dynamics in the TFIC by observing the time dependence of OTOC.
In experiment, we consider two different quenches as shown in the upper panels of Fig.~\ref{Fig1}(a):
(i) quenching from $g=0$ to $g=0.5$ and (ii) quenching from $g=0$ to $g=1.5$.
The whole evolution is divided into $M=12$ steps with fixed time increment $\tau=0.5$,
and the experimental results are shown in the upper panels of Fig.~\ref{Fig2}(a).
Only the real parts of the OTOC $F_\text{R}(t)$ are measured in experiment. In both quenches, $F_\text{R}(t)$ starts from $F_R(t=0)=1$ at $t=0$ and then decays due to the information spreading. Obviously, the long time behavior of the two cases are quite different. For $g=0.5$ where the Hamiltonian is in the ferromagnetic region, $F_\text{R}(t)$ oscillates as a function of time but is always positive.
In contrast, for $g=1.5$, ${F}_\text{R}(t)$ oscillates around zero \footnote{This result is to some extent different in comparison to the theoretical prediction in \cite{heyl2018detecting}, which states that $F_\text{R}(t)$ will reach zero. This discrepancy is mainly due to the small size of the simulated system;
see supplemental material~\cite{SupplementalMaterial}.}.

From the behaviour of $F_\text{R}(t)$, we can readily differentiate the dynamical ferromagnetic phases and paramagnetic phases. That is to say, there will be a DQPT in-between. For comparison, we measure the time evolution of the autocorrelation $\chi(t)=\langle \sigma_1^z(t)\sigma_1^z \rangle$
during the quench dynamics, with the experimental results shown in the lower panels of Fig.~\ref{Fig2}(a).
In theory, for quantum quench from the polarized state $\left|  \uparrow  \uparrow  \uparrow\ldots \right\rangle$, $\chi_\text{R}(t)=\langle \sigma_1^z(t)\rangle$ vanishes with time because the quantum system is heated by the quenching process. There is no long-range quantum order for one-dimensional models with short-range interaction at non-zero temperature. Indeed, we observe that $\chi_\text{R}(t)$ oscillates around zero in both quantum quenches [the lower panels of Fig.~\ref{Fig2}(a)], which indicates that the autocorrelation function cannot be used to signify the two different dynamical quantum phases, and thus cannot detect DQPTs. Therefore, we experimentally verify that the OTOC of the order parameter, as a four-point correlation function, can detect different dynamical quantum phases and the DQPTs, while the order parameter and the two-body correlation function can not.

\begin{figure}[tp]
\includegraphics[width=0.95\linewidth]{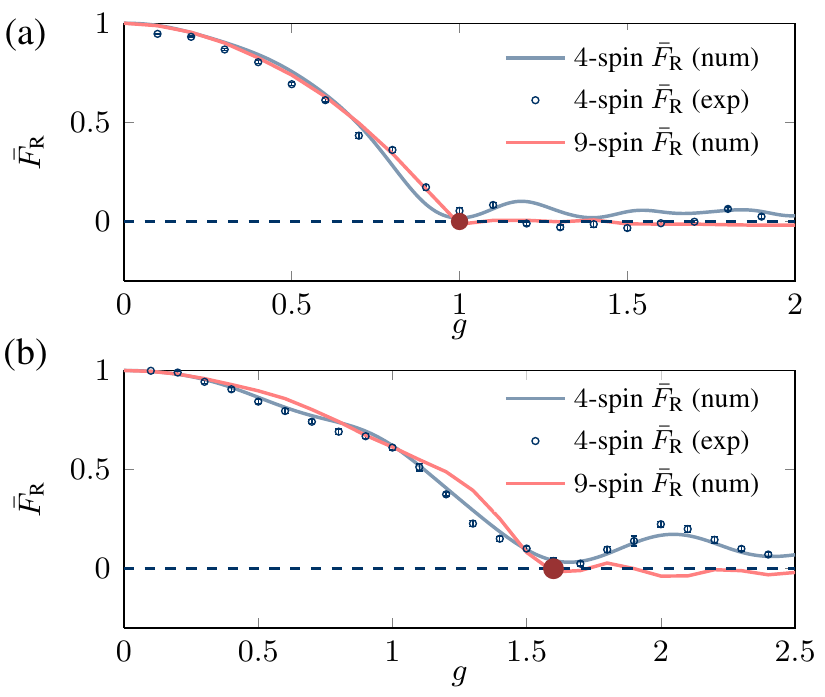}
\caption{Long-time averaged OTOC $\bar{F}_\text{R}$ as a function of the transverse field strength $g$ in
 (a) TFIC model and (b) ANNNI model.
 The blue dots and the solid lines represent the experimental data and numerical simulation results, respectively.
 The red dots represent the phase transition critical points of the two models
 between the ferromagnetic phase and the paramagnetic phase, respectively.
 The red lines are the simulation results with $N=9$ for comparison.}
\label{Fig3}
\end{figure}

Furthermore, we study how the long-time average of OTOCs $\bar{F}_\text{R}$
changes with the transverse field $g$.
We vary $g$ from $0.1$ to $1.9$ with increment $0.1$, and measure $\bar{F}_\text{R}$ during the time evolution of OTOCs.
The experimental results are shown in Fig.~\ref{Fig3}(a).
In the ferromagnetic phase, $\bar{F}_\text{R}$ is nonzero and eventually vanishes when approaching the equilibrium critical point $g=1$. In the paramagnetic phase, $\bar{F}_\text{R}$ stays zero.
This result confirms the validity of using the long-time averaged OTOC as an order parameter to detect DQPTs
as well as to locate the corresponding equilibrium quantum critical point.
The fluctuation beyond the critical point in the simulation result (blue dashed line) is due to the small size of the system.
We implement numerical simulations in larger systems ($N\geq9$), and find that the fluctuation is much lower (red solid line).

\emph{Non-integrable ANNNI mode..} --
We now turn to the non-integrable ANNNI model. Two different quantum quenches are investigated:
from $g=0$ to $g=0.5$ and from $g=0$ to $g=2.0$.
The dynamics is divided into $M=15$ segments with duration of each segment set as $\tau=0.2$.
The experimental OTOCs $F_\text{R}(t)$ are shown in the top panels of Fig.~\ref{Fig2}(b),
and the $\chi_\text{R}(t)$ are also measured for comparison, shown in the bottom panels of Fig.~\ref{Fig2}(b). From the results, it can be seen that, the two-body correlation function cannot distinguish different dynamical phases and the DQPTs, while the OTOC of the order parameter works fine. The long-time averaged OTOC as a function of the quenched parameter is experimentally observed in Fig.~\ref{Fig3}(b), where $g$ is varied from $0.1$ to $2.4$. Similar behaviors with the TFIC model are observed for the ANNNI model:
$\bar{F}_\text{R}$ takes a finite value at the ferromagnetic phase, gradually approaches zero when $g$ approaches the critical point $g_c\simeq1.6$,
and finally stays zero throughout the paramagnetic region. This is an evidence that OTOC in quench dynamics can be served as an order parameter to locate the equilibrium quantum critical point in the non-integrable cases, beating the autocorrelation function. The numerical simulation of
$N=9$ (red solid line in Fig.~\ref{Fig3}(b)) is also given.


\emph{Conclusion.} --
In this work, we present the first experimental observation of EQPTs and DQPTs from quench dynamics of OTOC in both integrable and nonintegrable Ising models on an NMR quantum simulator. To be concluded, both the order parameter and the ordinary two-body correlation function in quantum quench cannot be used as a probe to observe DQPTs. However, the OTOC, which is a four-point correlation function, can detect DQPTs. Therefore, our experiment unveils the important correlations between the OTOC and DQPTs. Moreover, our experiment demonstrates the feasibility of experimentally studying the EQPTs by performing a dynamical non-equilibrium measurement without carrying out the challenging initialization of the true many-body ground state. In addition to quantifying the information scrambling and diagnosing chaotic behavior of quantum many-body systems, our experiment establishes the OTOC as a faithful probe for DQPTs and EQPTs. While our work focuses on the short-range many-body systems, it would be interesting to investigate the relations among OTOCs, EQPTs and DQPTs in long-range many-body systems of one- and higher-dimensions.

\emph{Acknowledgments}.
This work is   supported by the National Key Research and Development Program of China (Grants No. 2019YFA0308100), the National Natural Science Foundation of China (Grants
No. 11605005, No. 11875159, No. 11905099, No. 11975117,   No. U1801661, and No. 11604220),
Science, Technology and Innovation Commission of Shenzhen
Municipality (Grants No. ZDSYS20170303165926217
and No. JCYJ20170412152620376), Guangdong Innovative
and Entrepreneurial Research Team Program (Grant No.
2016ZT06D348), Guangdong Basic and Applied Basic Research Foundation (Grant No. 2019A1515011383). B. B. W. also acknowledges the President's Fund of the Chinese University of Hong Kong, Shenzhen.

%

\clearpage
\appendix

\section{Appendix A. Experimental details}

\subsection{Initialization}
The main experimental procedure of the experiment is illustrated in Fig. 1(a) of the main text.
Firstly, starting from the thermal equilibrium state $\rho_\text{eq}\simeq\frac{1}{N}(\bm{I}-\sum_{i=1}^{4}\alpha_i\sigma_i^z)$ with $\alpha_i=\omega_i/k_\text{B}T$,
 we initialize the system to a Pseudo-pure state (PPS)~\cite{cory1997ensemble}
\begin{equation}\label{Eq1}
  \rho_0=\frac{1-\epsilon}{16}\bm{I}+\epsilon \vert 0000\rangle \langle 0000\vert.
\end{equation}
using the line-selective method~\cite{peng2001preparation},
here $\bm{I}$ is the $16\times16$ identity matrix,
$\epsilon\sim10^{-5}$ is the polarization of the nuclear system.
The PPS behaves similarly as the pure state $\vert 0000\rangle$ up to a scale factor $\epsilon$ on the readout signal.
The pulse sequence of the PPS preparation includes two line-selective pulses
together with two gradient fields to remove the undesired quantum coherence.
The first selective pulse aims to saturate the populations of all computational basis states except the first one,
and the second is designed to transfer the zero-order quantum coherence to the positions of the multi-order coherence.
Both line-selective pulses are realized using shaped pulses optimized by the GRAPE algorithm~\cite{khaneja2005optimal}, whose durations are $25$ ms and $20$ ms, respectively.

\subsection{Experimental measurement}
\emph{Ensemble readout.}--
The values of both OTOC and autocorrelation in the Fig. 2 of the main text are experimentally obtained by measuring
the expectation values of $\sigma_1^z$ of the final states.
Because the readout of an NMR experiment is an average over the nuclear spin ensemble,
the expectation value can be acquired by measuring the magnetization along $x$-direction
after applying a $\pi/2$ single spin rotation around the $y$-axis on the first nucleus in a single experiment.

\begin{figure}[thb]
  \renewcommand{\figurename}{Figure.}
  \centering
  \includegraphics[width=0.5\textwidth]{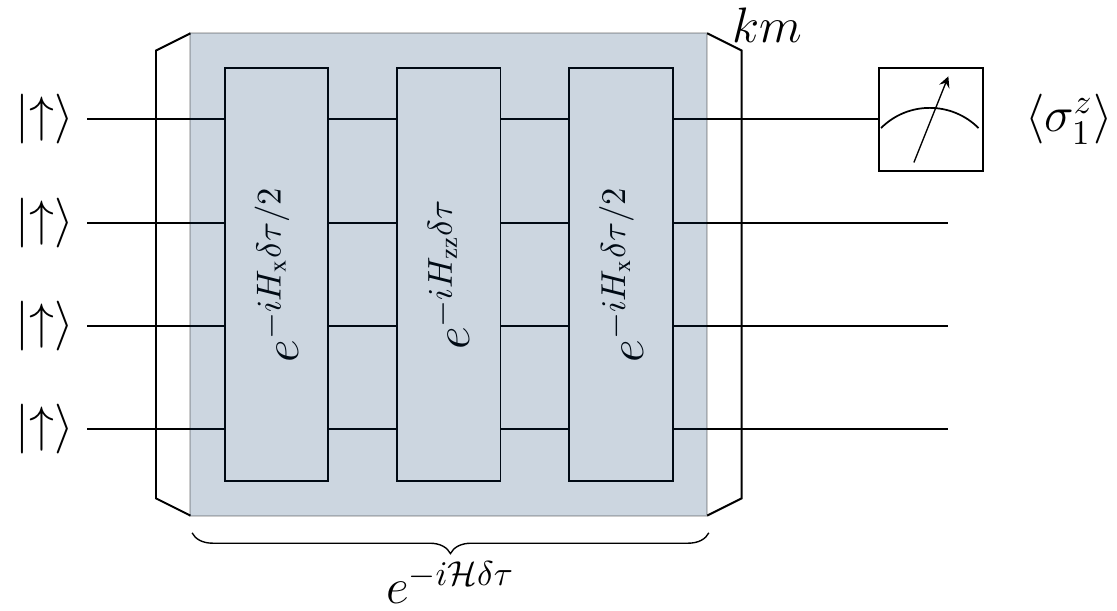}
  \caption{The quantum circuit used to measure the autocorrelation $\chi(t)$ in $4$-spin one-dimensional Ising models
   quenching from the fully polarized state $\vert\psi_0\rangle=\left|  \uparrow \uparrow \uparrow  \uparrow \right\rangle$. }\label{FigS1}
\end{figure}
\emph{Measurement of autocorrelation.}--
The quantum circuit to measure the autocorrelation is different from that of the OTOC.
In the quantum quench with initial state $\vert\psi_0\rangle=\vert0000\rangle$,
the autocorrelation can be written in the form of
\begin{equation}\label{Eq2}
 \chi(t)=\langle\Psi(t)\vert \sigma_1^z\vert \Psi(t)\rangle,
\end{equation}
where $\vert\Psi(t)\rangle=e^{-iHt}\vert\psi_0\rangle$.
Discrete the time duration as the same as the main text,
the quantum circuit of the measurement is shown in Figure.~\ref{FigS1},
the concrete form of $H_\text{x}$ and $H_\text{zz}$ can be found in the main text.
To reduce the experimental errors, we implement the unitary evolution $e^{-iHt}$  by a GRAPE pulse with
the time length of $40$ ms.

\section{Appendix B. Analyses of experimental Results}

\subsection{The influence of system size}
\begin{figure}
  \renewcommand{\figurename}{Figure.}
  \centering
  \includegraphics[width=0.49\textwidth]{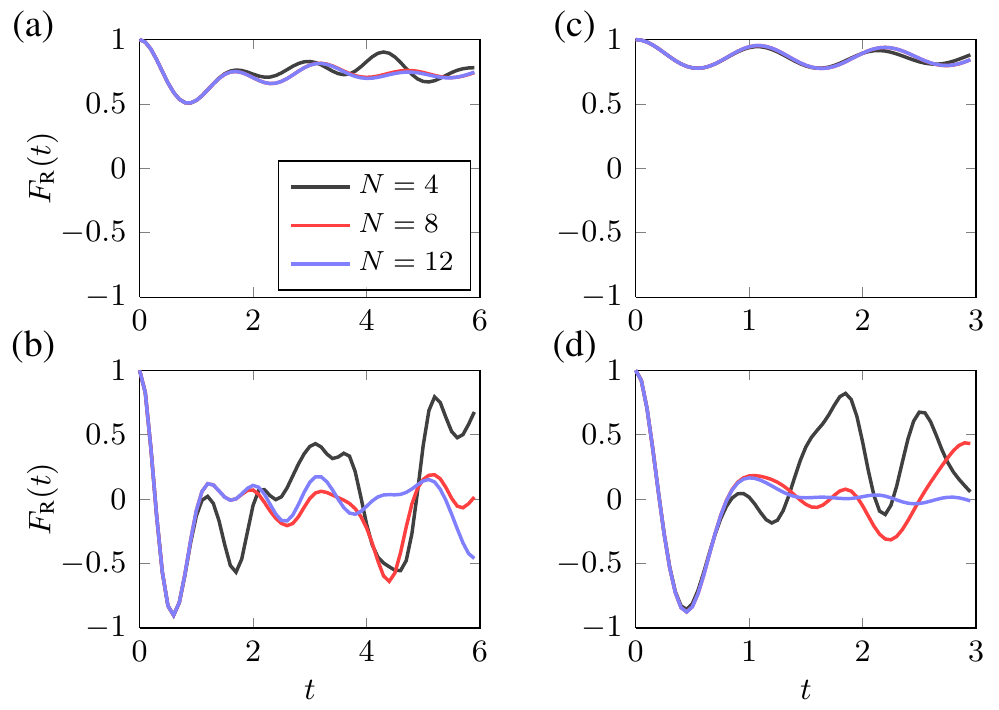}
  \caption{ The numerically simulated $F_\text{R}(t)$ following quench from the fully polarized
  the state is shown for the two Ising models with different spin numbers.
  The number of spins are chosen as $N=4$, $8$ and $12$, respectively.
  Time evolution of $F_\text{R}$ following quenches in the TFIC model from $g=0$ (a)  to  $g=0.5$ and  (b) to $g=1.5$;
  Time evolution of $F_\text{R}$ following quenches in the ANNNI model from $g=0$ (c) to $g=0.5$  and (d)  to $g=2.0$. }\label{FigS2}
\end{figure}

To analyze the difference between the experimental data in Fig. 2 of the main text and the theoretically predicted
results, we numerically simulate the time dependence of $F_\text{R}$ with different system sizes,
as shown in Figure.~\ref{FigS2}.
The simulation results in both TFIC and ANNNI models reveal that
the fluctuation of $F_\text{R}(t)$ of the second quench suppresses as the system size enlarges.
\subsection{Error analysis}
We use the average abstract deviation
$\delta=\frac{1}{K}\vert x^i_\text{exp}-x^i_\text{th}\vert$ to estimate the error in Fig. 2 and Fig. 3 of the main text
and their average error are $0.081$ and $0.018$, respectively,
here $K$ is the number of the data points,  $x^i_\text{exp}$ and $x^i_\text{th}$ represent the $i$-th experimental and theoretical values.
It shows that the error of $\bar{F}_\text{R}$ is much smaller than that of $F_\text{R}$, which is mainly due to that
the random error term is reduced by multiple measurements.
These experimental errors may be caused by the imperfection of the initial state,
the inaccuracy of the GRAPE pulse, the effect of decoherence during the experimental time
and the sampling error.
Considering that the duration of quantum circuit in experiments is much shorter than the relaxation time,
the effect of decoherence can be ignored.
Besides, the fidelities of the GRAPE pulses are all above $99.5\%$, which will result in an error of about $0.3\%$.
In the error analysis, we mainly consider the effect of the imperfect initial states and the readout error.
The imperfect of the initial state with a fidelity of $95\%$  adds an average error of $2.5\%$ and $3.3\%$
to the experimental results of TFIC and ANNNI, respectively.
The readout error is coursed by the white noise of the NMR spectra, which is about $2\%$.
\begin{figure}
  \renewcommand{\figurename}{Figure.}
  \centering
  \includegraphics[width=0.5\textwidth]{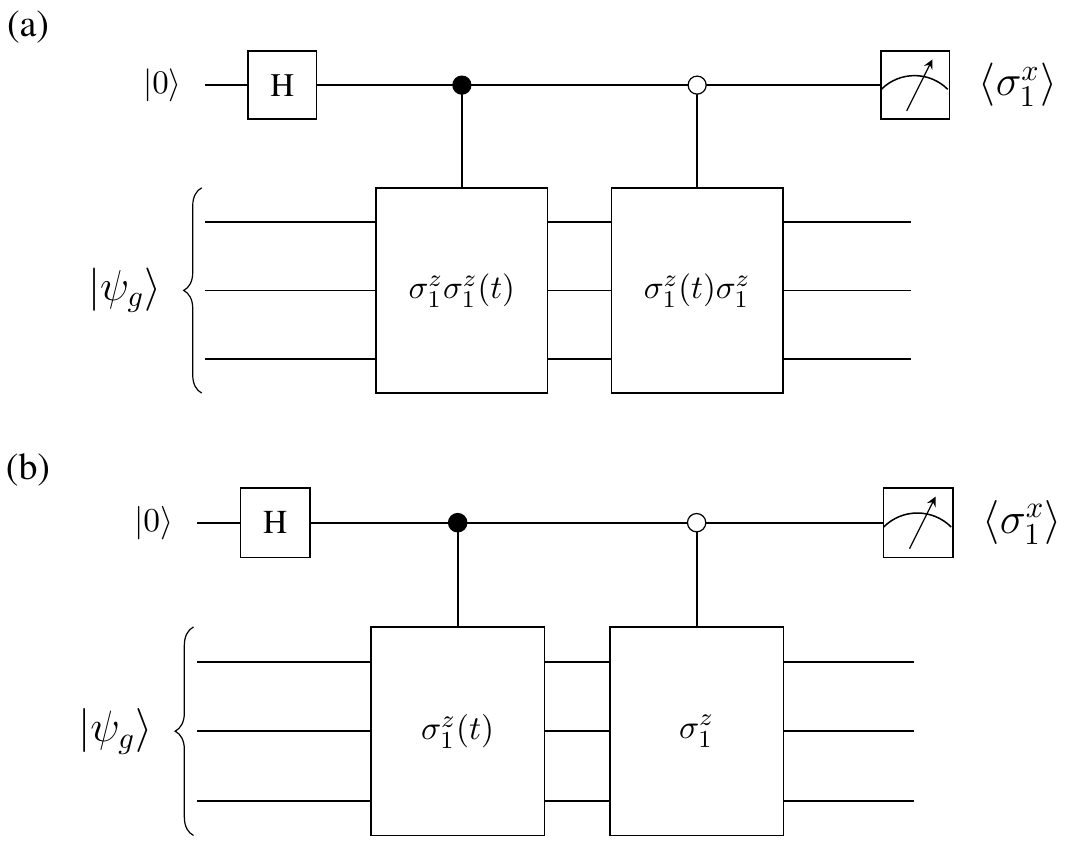}
  \caption{The experimental quantum circuits of the measurements of in the equilibrium dynamics of the $3$-spin Ising models of the equilibrium quantum phase transition using a four-qubit quantum simulator.
  (a) The circuit to measure the real part of (a) OTOC $F_\text{R}(t)$.
  (b) The circuit to measure the real part of autocorrelation $\chi_\text{R}(t)$. }\label{FigS3}
\end{figure}

\begin{figure*}[thb]
  \renewcommand{\figurename}{Figure.}
  \centering
  \includegraphics[width=1.0\textwidth]{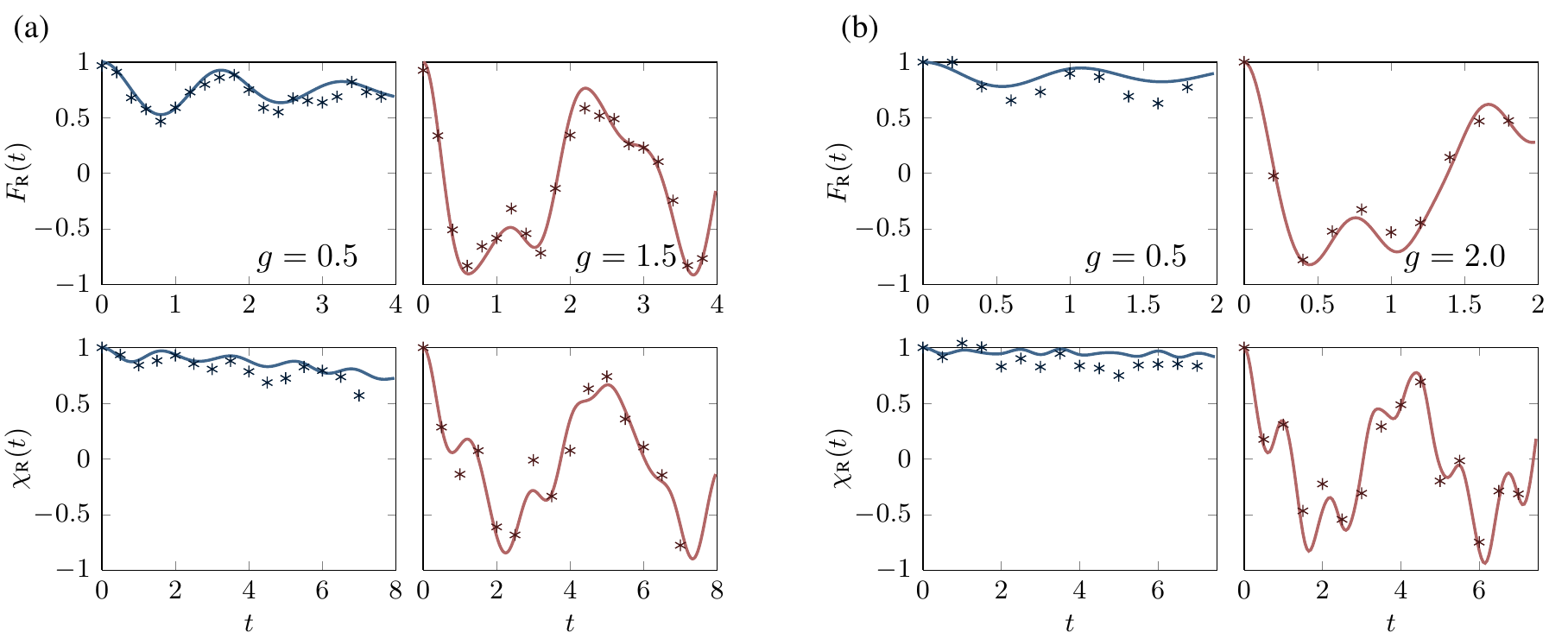}
  \caption{The time dependence of $F_\text{R}$ and $\chi_\text{R}$ of the equilibrium quantum dynamics for the $3$-spin (a) TFIC and (b) ANNNI models.
  The upper panels are the time evolutions of $F_\text{R}$, and the lower panels are the time evolutions of $\chi _\text{R}$. Stars are experimental results, and the solid lines are numerical simulation results with the experimental parameters.}\label{FigS4}
\end{figure*}

\begin{figure}[h]
  \renewcommand{\figurename}{Figure.}
  \centering
  \includegraphics[width=0.5\textwidth]{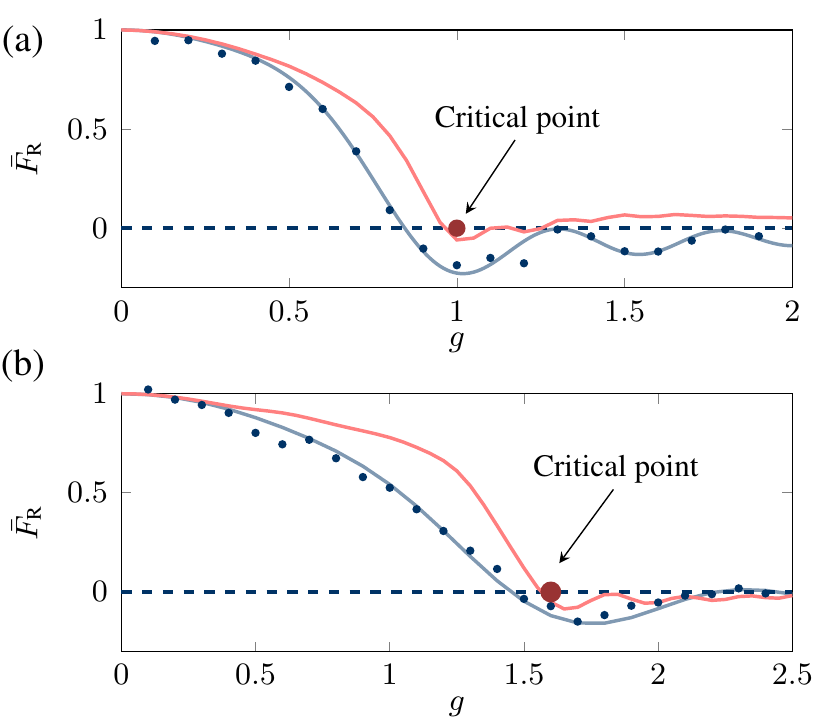}
  \caption{The long-time averaged OTOC as a function of $g$ for the (a) TFIC and (b) ANNNI model.
  The dots are the experimental data and the solid lines are the numerical simulation results with the same parameters as the experiment, the red solid lines are the numerical simulation results of $8$-spin Ising models.
  The red dots are the theoretical quantum critical points.}\label{FigS5}
\end{figure}

\section{Appendix C. Observation of quantum phase transition via OTOC in equilibrium dynamics}
\subsection{Experimental Implementation}
We also experimentally demonstrate that the OTOC of the equilibrium dynamics can be used to detect quantum phase transition and the critical points in both integrable and non-integrable quantum systems.
The initial state of equilibrium dynamics is the ground state of the Hamiltonian $H$, i.e., $\vert\psi_0\rangle=\vert\psi_g\rangle$.
As same as the experiments of quench dynamics in the main text, we investigate the TFIC and ANNNI models in the equilibrium case.
Comparing with the quench experiment, the experimental measurement of the OTOCs in an equilibrium case is more complicated.
First, we have to prepare the ground state of the Hamiltonian $H$ with different $g$,
which is much more difficult than the fully polarized state;
Second, for $\vert \psi_g\rangle$ is not an eigenstate of the local operator $\sigma_1^z$,
we cannot rewrite $F(t)$ and $\chi(t)$ as in the quantum quench case so that they cannot be measured with quantum circuit shown in Fig. 1(c) of the main text or the quantum circuit in Figure.~\ref{FigS1}.
In the following, we will explain the experimental details and show the experimental results of the equilibrium cases.

\textit{Quantum circuits}.--
We write the OTOC in the equilibrium dynamics in the form of $F(t)=\langle\psi_1(t)\vert \psi_2(t)\rangle$,
with $\vert\psi_1(t)\rangle=\sigma_1^z\sigma_1^z(t)\vert\psi_g\rangle$ and $\vert\psi_2(t)\rangle=\sigma_1^z(t)\sigma_1^z\vert\psi_g\rangle$.
Such kind of state overlap can be measured by introducing an ancillary qubit.
We use the same $4$-spin nuclear magnetic system as the quantum simulator
and study the OTOC behaviors of the $3$-spin Ising models except for an ancillary qubit.
The quantum system is first initialized into a product state $\vert0\rangle\otimes\vert\psi_g\rangle$.
Then a Hadamard gate is applied on the ancillary qubit
so as to get a state $\frac{1}{\sqrt{2}}(\vert0\rangle+\vert1\rangle)\otimes \vert\psi_g\rangle$.
After that the system is evolved with two control gates in the form of $U_1=\vert1\rangle\langle1\vert\otimes\sigma_1^z\sigma_1^z(t)$
and $U_2=\vert0\rangle\langle0\vert\otimes\sigma_1^z(t)\sigma_1^z$.
The real part of OTOC can be read out by measuring the expectation value of $\sigma_x^1$.
The quantum circuit is shown in Figure.~\ref{FigS3}(a).

We also experimentally study the time evolution of autocorrelation $\chi(t)=\langle\sigma_1^z(t)\sigma_1^z \rangle$ in  the equilibrium dynamics.
The quantum circuit is similar to that of the OTOC as shown in Figure.~\ref{FigS3}(b).
The only difference is in the two control gates, which are in the form of  $U_1=\vert1\rangle\langle1\vert\otimes\sigma_1^z(t)$
and $U_2=\vert0\rangle\langle0\vert\otimes\sigma_1^z$, respectively.

\textit{Preparation of the ground states}.--
The initial state of the equilibrium dynamics is the ground state of the Hamiltonian $H$,
i.e., $\vert\psi_0\rangle=\vert\psi_g\rangle$.
It is usually difficult to prepare the ground state in experiment, because the concrete form of $\vert\psi_g\rangle$ depends on the value of $g$ and is always highly entangled.
Fortunately, in the case of small system, the ground state can be exactly solved, we can search a unitary evolution $U_g$
which satisfies $\vert\psi_g\rangle=U_g\vert00\cdots0\rangle$, such that the ground state $\vert\psi_g\rangle$ can be
prepared from $\vert00\cdots0\rangle$ directly.
In the case of large system, one may have to design an adiabatic passage so as to obtain the ground state for a particular $g$.

In principle, all the unitary operations can be realized with the well-designed pulse sequence
composed of the universal quantum gates, but it always results in tremendous experimental errors.
In the experiments, we use a single shaped pulse searched by GRAPE algorithm to realize the unitary evolution
$U_2U_1HU_g$, the time length and the theoretical fidelity of the shaped pulse are $40$ ms and $99.5\%$, respectively.

\subsection{Experimental Results and Discussion}
The experimental results of the equilibrium dynamics are shown in Figure.~\ref{FigS4} and Figure.~\ref{FigS5}.
Firstly, we observe the time dependence of OTOC and autocorrelation in equilibrium dynamics with
the time evolving Hamiltonian in different regions for both TFIC and ANNNI models, with experimental results shown in Figure.~\ref{FigS4}.
We find that $F_\text{R}(t)$ behaves analogously as that of the quantum quench in both Ising models:
in the case where Hamiltonian locates in the ferromagnetic region, $F_\text{R}$ maintains a positive value as the time goes by;
where in the region of paramagnetic phase, $F_\text{R}$ diminishes to a small value nearby zero.
This means that the time dependence of OTOC can be used to characterize the kind of Hamiltonian.
Besides, the behavior of $\chi(t)$ also changes with the Hamiltonian:
when the Hamiltonian is ferromagnetic, $\chi_\text{R}(t)$ oscillates around a positive value;
while for the paramagnetic Hamiltonian, $\chi_\text{R}(t)$ diminishes to a negative value and then revives.
This illustrates that $\chi(t)$ can be used to differentiate the kind of the Hamiltonian in the equilibrium dynamics,
which is different with the non-equilibrium dynamics.

The experimental long-time averaged OTOCs of the two Ising models are shown in Figure.~\ref{FigS5}:
$\bar{F}_\text{R}$ gradually decreases to zero when approaching the corresponding theoretical critical points at $g_c=1.0$
and $g_c=1.6$ for the TFIC and ANNNI models,
and oscillates around a small value near zero in the whole paramagnetic phase region.
However, there are obvious distinctions between the theoretical critical points and the experimental ones,
which is mainly due to the small size of the Ising models simulated in the experiments.
We also numerically calculate $\bar{F}_\text{R}$ in larger Ising systems with $N=8$ spins,
which are shown with the red solid lines.
It is an obvious evidence that the difference between experimental data and the theoretical predictions can be removed as the system grows larger.

%

\end{document}